\documentclass[useAMS,usenatbib]{mn2e}
\pdfoutput=1

\usepackage{graphicx}
\usepackage{amsfonts,amssymb}
\usepackage{times}
\renewcommand{\vec}[1]{\mbox{\boldmath $#1$}}
\def\Om{\Omega}

\def\q{\qquad}
\def\beg{\begin{eqnarray}}
\def\ende{\end{eqnarray}}
\def\gsim{\lower.4ex\hbox{$\;\buildrel >\over{\scriptstyle\sim}\;$}} 
\def\lsim{\lower.4ex\hbox{$\;\buildrel <\over{\scriptstyle\sim}\;$}}

\renewcommand{\textrm} [1] {\rm #1} 
\renewcommand{\textit} [1] {\it #1}
\renewcommand{\textbf} [1] {\bf #1}

\title[Helicity and alpha-effect by current-driven instabilities of helical magnetic fields]
{Helicity and alpha-effect by current-driven instabilities of helical magnetic fields}
 \author[M. Gellert, G. R\"udiger and R. Hollerbach]
{
M. Gellert$^1$\thanks{E-mail: mgellert@aip.de}, G. R\"udiger$^{1,2}$, R. Hollerbach$^3$\\
$^1$Astrophysikalisches Institut Potsdam,
         An der Sternwarte 16, D-14482 Potsdam, Germany\\  $^2$Forschungszentrum Dresden Rossendorf, P.O. Box 510119, D-01314 Dresden, Germany
\\
$^3$Department of Applied Mathematics, University of 
Leeds, Leeds, LS2 9JT, UK}

\begin{document}

\date{Accepted . Received ; in original form }
\pagerange{\pageref{firstpage}--\pageref{lastpage}} \pubyear{2011}
\maketitle
\label{firstpage}


\begin{abstract}
Helical magnetic background fields with adjustable pitch angle are imposed on a conducting fluid in a differentially rotating cylindrical container.  The small-scale kinetic and current helicities are calculated for various field geometries,
and shown to have the opposite sign as the helicity of the large-scale field.  These helicities and also the corresponding $\alpha$-effect scale with the current helicity of the background field. The $\alpha$-tensor is highly anisotropic as the components $\alpha_{\phi\phi}$ and $\alpha_{zz}$ have opposite signs. The  amplitudes of the azimuthal $\alpha$-effect computed with the cylindrical 3D MHD code are so small that the operation of an $\alpha\Omega$ dynamo on the basis of the current-driven, kink-type instabilities of toroidal fields is highly questionable. In any case the low value of the $\alpha$-effect would lead to very long growth times of a dynamo in the radiation zone of the Sun and early-type stars of the order of mega-years.
\end{abstract}

\section{Introduction}
No hydromagnetic dynamo can exist driven only by differential rotation \citep{E46}, but it is known that such dynamos can exist  if the turbulence is helical in the sense that its kinetic helicity
\begin{equation}
{\cal H}_{\rm kin}= \langle \vec{u} \cdot{\rm curl}\,\vec{u}\rangle
\label{1}
\end{equation}
and/or its current helicity
\begin{equation}
{\cal H}_{\rm curr}= \frac{1}{\mu_0\rho}\ \langle \vec{b}\cdot {\rm curl}\, \vec{b}\rangle
\label{2}
\end{equation}
do not vanish. Here $\vec{u}$ and $\vec{b}$ are the fluctuating parts of the flow  $\vec{U}$ and magnetic field $\vec{B}$. This condition of non-vanishing helicity is clearly fulfilled if the turbulence is rotating and stratified. In such turbulence a pseudo-scalar exists which allows the pseudo-scalars (\ref{1}) and (\ref{2}) to take finite values. The same is true for linear shear flows where the stratified turbulence in the presence of the shear also can form a kinetic helicity (see R\"udiger \& Kitchatinov 2006). The simplest pseudo-scalar is the scalar product $\vec{g}\cdot \vec{\Omega}$ with $\vec{g}$ as the gradient vector of the turbulence (or the fluid density) and $\vec{\Omega}$ the rotation vector. In spheres the gradient vector $\vec{g}$ is mainly radial so the pseudo-scalar $\vec{g}\cdot \vec{\Omega}$ has opposite signs in the two hemispheres, and vanishes at the equator. Because of the close relationship of the helicity to the $\alpha$-effect in the mean-field electrodynamics of turbulent media,
\begin{equation}
\langle \vec{u}\times \vec{b}\rangle= \alpha \vec{B}_0 + \dots 
\label{3}
\end{equation}
(the dots represent higher derivatives of $\vec{B}_0$) the above-mentioned sign rules are also the sign rules of the $\alpha$-effect i.e. $\alpha \propto \vec{g} \cdot \vec{\Omega}$.

This sort of $\alpha$-effect only exists for inhomogeneous turbulence. In planetary cores, however, and also in laboratory experiments the only inhomogeneities result from boundary conditions, as the density gradients are negligible. One can show that under the presence of magnetic background fields other inhomogeneities also form pseudo-scalars and, as a consequence, lead to new mechanisms for an $\alpha$-effect (e.g. Gellert, R\"udiger \& Elstner 2008). In the present study we demonstrate that instabilities due to inhomogeneous magnetic fields also lead to finite values of the helicities (\ref{1}) and (\ref{2}), and in accord with (\ref{3}) also to finite values of $\alpha$. Indeed, in the presence of electric currents the simplest existing pseudo-scalar is $\vec{B}_0 \cdot {\rm curl}\, \vec{B}_0$ which does not vanish for helical field geometries. We show that for such background fields the small-scale helicities obtain final values with the opposite sign as the helicity of the background field.

According to the Rayleigh criterion, in the absence of MHD effects an ideal flow is stable against axisymmetric perturbations whenever the
specific angular momentum increases outward
\beg
\frac{{\rm{d}}}{{\rm{d}}R}(R^2\Omega)^2 > 0,
\label{ray}
\ende
where $\Omega$ is the angular velocity, and ($R$, $\phi$, $z$) are
cylindrical coordinates in a right-handed system.  In the presence of an azimuthal magnetic field $B_\phi$, this criterion is modified to
\beg
\frac{1}{R^3}\frac{{\rm{d}}}{{\rm{d}}R}(R^2\Omega)^2-\frac{R}{\mu_0\rho}
\frac{{\rm{d}}}{{\rm{d}}R}\left( \frac{B_{0,\phi}}{R} \right)^2 > 0,
\label{mich}
\ende
where $\mu_0$ is the permeability and $\rho$ the density \citep{M54}.
Note also that this criterion is both necessary and sufficient for
(axisymmetric) stability. In particular, {\em all} ideal flows can thus
be destabilized, by azimuthal magnetic fields with the right profiles (steeply increasing outwards) and amplitudes.

On the other hand, for nonaxisymmetric modes one has
\beg
\frac{{\rm{d}}}{{\rm{d}}R}( R B_{0,\phi}^2) < 0
\label{tay}
\ende
as the necessary and sufficient condition for stability of an ideal fluid at rest \citep{T73}. Again, outwardly increasing fields are thus unstable. If  (\ref{tay}) is violated, the most unstable mode has azimuthal wave numbers of $m=\pm 1$. 

\section{Equations}
We are interested in the  stability of the background field
$\vec{B_0}= (0, B_\phi(R), B_0)$, with $B_0=\rm const$, and the flow
$\vec{u_0}= (0,R\Omega(R), 0)$.
The governing equations for the flow $\vec{U}$ and the field $\vec{B}$ are
\begin{eqnarray}
\frac{\partial \vec{U}}{\partial t} + (\vec{U}\cdot\nabla)\vec{U}
 =
-\frac{1}{\rho} \nabla p + 
\nu \Delta \vec{U} + \nonumber\\
\quad\quad +\frac{1}{\mu_0\rho}{\textrm{curl}}\ \vec{B} \times \vec{B},
\label{mhd}
\end{eqnarray}
\begin{eqnarray}
\frac{\partial \vec{B}}{\partial t}= {\textrm{curl}}\ (\vec{U} \times \vec{B})+  \eta \Delta\vec{B},
\label{mhd1}
\end{eqnarray}
and
$
{\textrm{div}}\ \vec{U} = {\textrm{div}}\ \vec{B} = 0,
$
where $p$ is the pressure, $\nu$ the kinematic viscosity and $\eta$
the magnetic diffusivity. Their ratio is the magnetic Prandtl number
\beg
{\rm Pm} =\frac{\nu}{\eta}.
\label{Pm}
\ende
From now on we drop the subscripts from the large-scale values so that the total flow is $\vec{U}+\vec{u}$ and the total field is $\vec{B}+\vec{b}$.
The  stationary background solution is 
\beg
\Omega=a_\Omega +\frac{b_\Omega}{R^2},  \ \ \ \ \
B_\phi=a_B R+\frac{b_B}{R},
\label{basic}
\ende
where $a_\Omega$, $b_\Omega$, $a_B$ and $b_B$ are constants defined by 
\beg
a_\Omega=\Omega_{\rm{in}}\frac{ \mu_\Omega-{\hat\eta}^2}{1-{\hat\eta}^2}, \q
b_\Omega=\Omega_{\rm{in}} R_{\rm{in}}^2 \frac{1-\mu_\Omega}{1-{\hat\eta}^2},
\nonumber \\
a_B=\frac{B_{\rm{in}}}{R _{\rm{in}}}\frac{\hat \eta
( \mu_B - \hat \eta)}{1- \hat \eta^2},  \q
b_B=B_{\rm{in}}R _{\rm{in}}\frac{1-\mu_B \hat\eta}
{1-\hat \eta^2},
\label{ab}
\ende
with
\begin{equation}
\hat\eta=\frac{R_{\rm{in}}}{R_{\rm{out}}}, \; \; \;
\mu_\Omega=\frac{\Omega_{\rm{out}}}{\Omega_{\rm{in}}},  \; \; \;
\mu_B=\frac{B_{\rm{out}}}{B_{\rm{in}}}.
\label{mu}
\end{equation}
$R_{\rm{in}}$ and $R_{\rm{out}}$ are the radii of the inner and outer
cylinders, $\Omega_{\rm{in}}$ and $\Omega_{\rm{out}}$ are their rotation
rates, and $B_{\rm{in}}$ and $B_{\rm{out}}$ are the azimuthal magnetic fields
at the inner and outer cylinders. A  field of the form $b_B/R$ is generated by running an axial current only through the inner region $R<R_{\rm{in}}$, whereas a
field of the form $a_B R$ is generated by running a uniform axial current
through the entire region $R<R_{\rm{out}}$, including the fluid.  

Given the $z$-component of the electric current, ${\rm curl}_z\,\vec{B}= 2a_B$,
one finds for the current helicity of the background field
$
\vec{B}\cdot{\rm curl}\, \vec{B}  =2 a_B B_0,
$
which may be positive, negative, or zero.

The inner value $B_{\rm in}$ is normalized by the uniform vertical field,
i.e.
\begin{equation}
\beta =\frac{B_{\rm in}}{B_0}.
\label{beta}
\end{equation}
For our standard profile $\mu_B=1$ we have
$
\vec{B}_0\cdot{\rm curl}\,\vec{B}_0=2\beta \   B_0^2/\rho R_{\rm in}
$
for the helicity of the background field. For fixed toroidal field amplitude this quantity scales as $\beta^{-1}$:
\begin{equation}
 \vec{B}_0 \cdot {\rm curl}\, \vec{B}_0=\frac{2}{3 \beta}\ \frac{B_{\rm in}^2}{R_{\rm in}}.
\label{heli}
\end{equation}
The sign of $\beta$  determines the sign of the current helicity.  If the toroidal field is due to the interaction of a poloidal field with a differential rotation with negative shear then $\beta$ is negative and vice versa.  Interchanging $\pm\beta$ simply interchanges left and right spirals, $ m\to -m$.

As usual, the toroidal field amplitude is measured by the  Hartmann number and the global rotation by the Reynolds number, i.e.
\beg
{\rm Ha} = \frac{B_{\rm in} D}{\sqrt{\mu_0 \rho \nu \eta}}, \ \ \ \ \ \ \ \ \ \ \ \ \ \ \ \ \ \ \ \ \ \   {\rm Re}=\frac{\Om_{\rm in}  D^2}{\nu}.
\label{Ha}
\ende
$D=R_{\rm out} - R_{\rm in}$ is used as the unit of length,
$\nu/D$ as the unit of velocity and $B_{\rm in}$ as the unit of the azimuthal
fields. Frequencies, including the rotation $\Om$, are normalized with the inner rotation rate $\Om_{\rm in}$. The Lundquist number $\rm S$ is defined by $\rm S= \sqrt{Pm}\ Ha$.  The magnetic-diffusion frequency is $\omega_\eta= \eta/D^2$ and then the Alfv\'en frequency $\Omega_{\rm A}={\rm S} \ \omega_\eta$ is 
\beg
\Omega_{\rm A} =\frac{B^2_{\rm in}}{\mu_0\rho D}.
\label{OA}
\ende
Throughout the whole paper numerical values of helicities are given in units of $\Omega^2_{\rm A} D$. In this notation the helicity (\ref{heli}) of the background field can be written as 
\beg
\frac{1}{\mu_0\rho}    \vec{B}_0 \cdot {\rm curl}\, \vec{B}_0 =\frac{2}{3\beta}   \Omega_{\rm A}^2 D \frac{D}{R_{\rm in}}\simeq  \frac{\Omega_{\rm A}^2 D}{\beta},
\label{heli2}
\ende

The boundary conditions associated with the perturbation equations are no-slip
for $\vec{u}$ and perfectly conducting for $\vec{b}$, at both $R=R_{\rm{in}}$ and $R=R_{\rm{out}}$, where we fix $R_{\rm out}=2 R_{\rm in}$, i.e. $\hat\eta=0.5$. The computational domain is periodic in $z$. The nonlinear MHD code used for the solution of Eqs. (\ref{mhd}) and (\ref{mhd1}) has been described in detail by
\citet{Ge07} (see also Fournier et al. 2004, 2005).

\section{Results}
\begin{figure}
\includegraphics[width=8.5cm,height=6cm]{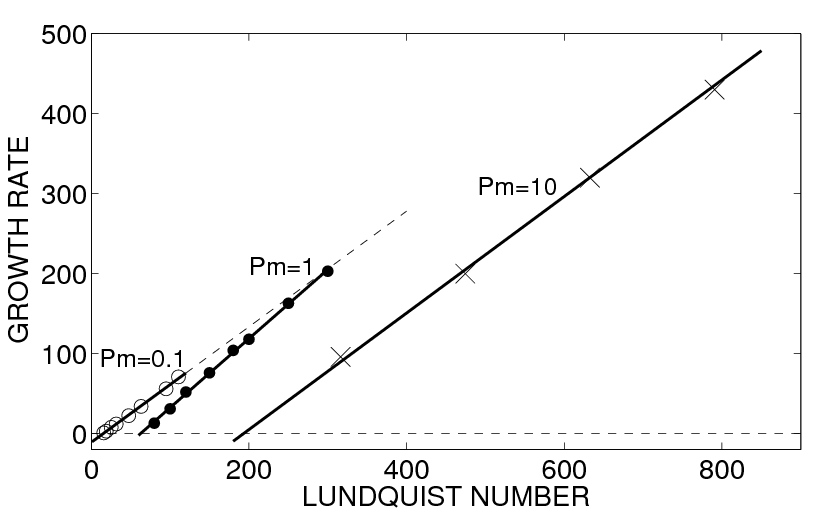}
\caption{\label{growth}
Growth rate curves for various magnetic Prandtl numbers, for stationary
cylinders and a purely toroidal field, with $\mu_B=1$. The growth rate and
the magnetic field are normalized with the magnetic-dissipation frequency
$\omega_\eta$.}
\end{figure}

Fig.~\ref{growth} shows the growth rates for a purely toroidal field
($\beta=\infty$), and no rotation.  We see that beyond the critical
Lundquist number, the growth rate is essentially linear, i.e.
\beg
\omega_{\rm gr} \propto \Omega_{\rm A},
\label{ogr}
\ende
where $\rm Pm=1$ has the steepest slope, and is thus more unstable than
both $\rm Pm<1$ and $\rm Pm>1$.

The azimuthal wavenumber of the modes shown in Fig. \ref{growth} is $m=\pm1$.
For a purely toroidal field, $\pm m$, corresponding to left- and right-handed
spirals, are degenerate, and necessarily have exactly the same growth rate
curves.  See also \citet{HTR}, who obtained the same effect in
magnetorotational instabilities, and \citet{RueKi11}, who consider
instabilities of toroidal fields in spheres.
\begin{figure}
\mbox{
\includegraphics[width=4.5cm,height=6cm]{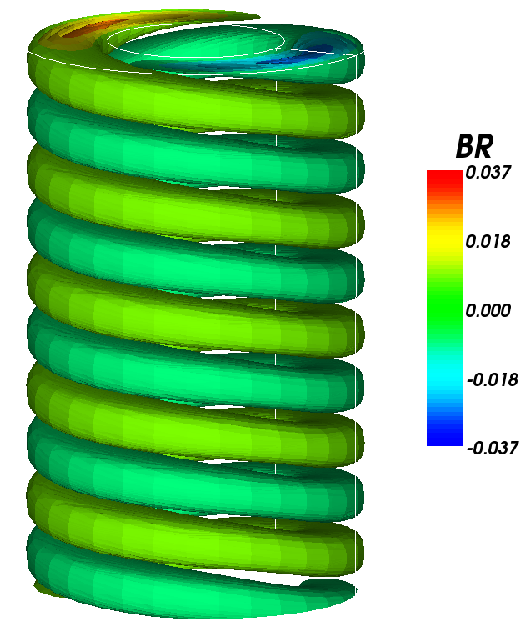}
\includegraphics[width=4.5cm,height=6cm]{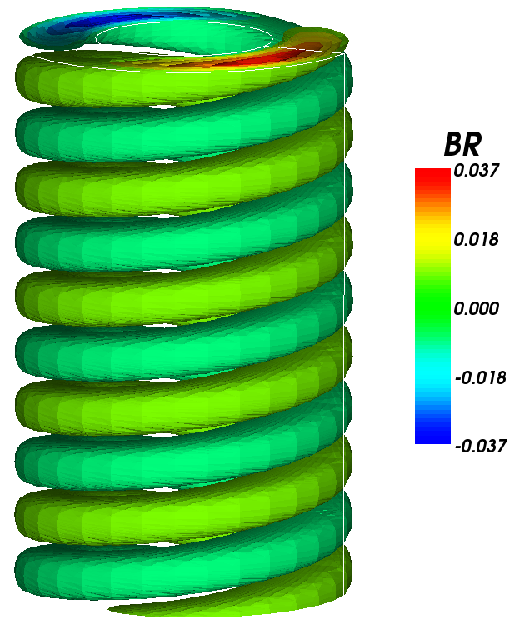}}
\caption{\label{heltor}
The left and right spirals that can be excited by a purely toroidal basic
state.  Apart from having the opposite handedness, the two modes are exactly
equivalent; their kinetic helicity is $\pm6.0\cdot 10^{-4}$ and their
current helicity is $\pm3.5\cdot 10^{-3}$ (both in units of
$\Omega^2_{\rm A} D$), where the left spiral is positive and the right
spiral negative.  These modes do not drift in the azimuthal direction.
The parameter values are $\beta=\infty$, $\mu_B=1$, $\rm Re=0$, $\rm Ha=200$,
and $\rm Pm=1$.}
\end{figure}

We next consider the nonlinear equilibration of these modes.  As Fig.
\ref{heltor} shows, even though $m=\pm1$ are degenerate, the equilibrated
solutions do not consist of equal mixtures of both modes.  Instead, either
the left or the right mode wins out, and completely suppresses the other.
Which mode one obtains depends on the precise initial conditions.  If these
already favor one mode, then (not surprisingly) that one wins, but if the
initial condition is evenly balanced between the two modes, it is ultimately
just numerical noise that determines which mode wins.  Eventually though one
mode always wins; the solution consisting of an equal mixture of both is
unstable.

Spontaneous parity-breaking bifurcations of this type are well known in
classical, non-magnetic Taylor-Couette flow (e.g. Hoffmann et al. 2009, Altmeier et al. 2010
and reference therein), but are almost unknown in magnetohydrodynamic
problems.  To the best of our knowledge, the only other example is in the
very recent work by \citet{Axel}.  Given the importance of helicity in
mean-field dynamics, any effect that generates helicity from an underlying
basic state without helicity could be significant.

\begin{figure}
\includegraphics[width=8.5cm,height=6cm]{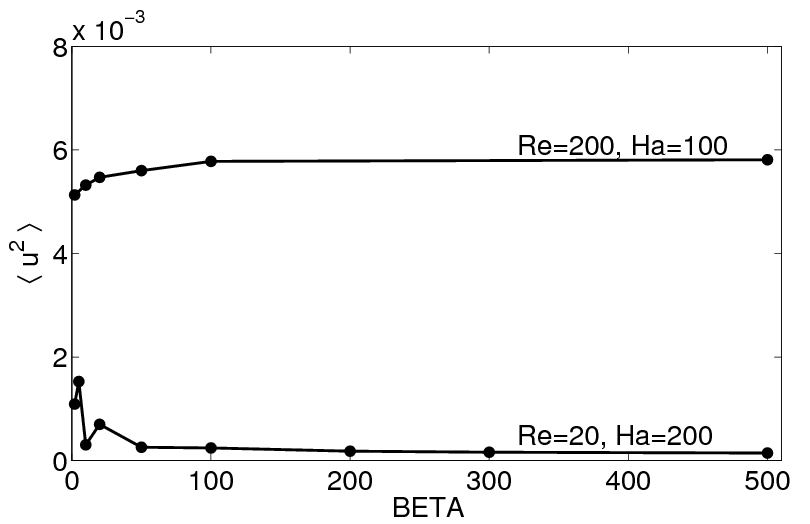}
\includegraphics[width=8.5cm,height=6cm]{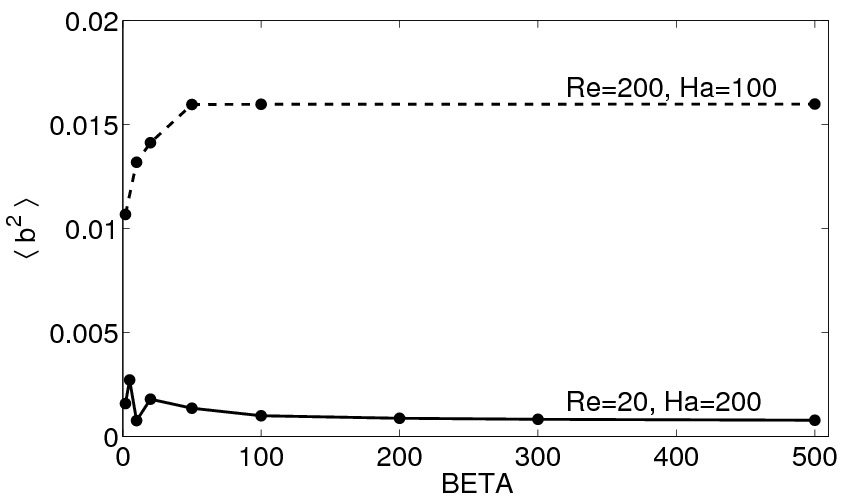}
\caption{\label{energy}
The turbulence intensities for flow (top) and field (bottom) fluctuations in
units of $D^2\Omega^2_{\rm A}$. Note that once $\beta$ exceeds $\sim100$ it
has virtually no further influence. $\mu_B=1$, $\mu_\Omega=0.5$, $\rm Pm=1$.}
\end{figure}

We next present two series of solutions where $\vec{B}_0$ includes an axial
component ($\beta<\infty$). In contrast to Figs. \ref{growth} and \ref{heltor},
we also include a differential rotation here.  The profiles of the basic state
field and flow are fixed at $\mu_B=1$ and $\mu_\Omega=0.5$. Their amplitudes
are $\rm Ha=100$ and $\rm Re=200$ for the first series, and $\rm Ha=200$,
$\rm Re=20$ for the second. The first series is thus rotationally dominated
($\Omega>\Omega_{\rm A}$), whereas the second is magnetically dominated
($\Omega_{\rm A}>\Omega$). The astrophysically relevant case is rotationally
dominated, which is {\emph not} the classical realization of the Tayler
instability. We have called this instability the Azimuthal Magnetorotational Instability (AMRI, see Hollerbach, Teeluck \& R\"udiger 2009).

For both series of runs Fig. \ref{energy} shows the kinetic and magnetic
turbulence intensities $\langle \vec{u}^2\rangle$ and $\langle\vec{b}^2\rangle$.
For sufficiently large $\beta$ its influence is very small; the axial
component of ${\vec B}_0$ is then so weak that it has no further influence.
This is not true for small $\beta$, where the axial field starts to dominate.
For $\beta<1$ the kink-instability is strongly stabilized \citet{Rshuel11} and
the resulting energies of the perturbations are reduced.

\begin{figure}
\includegraphics[width=8.5cm,height=6cm]{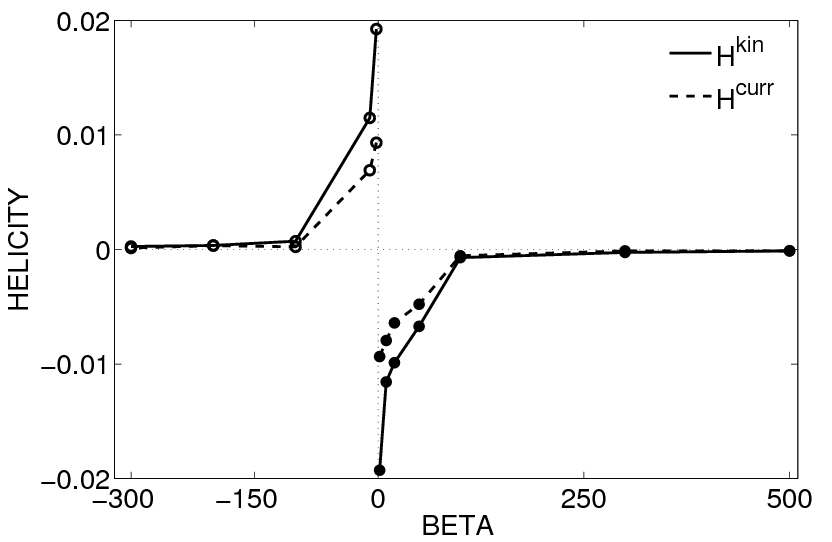}
\includegraphics[width=8.5cm,height=6cm]{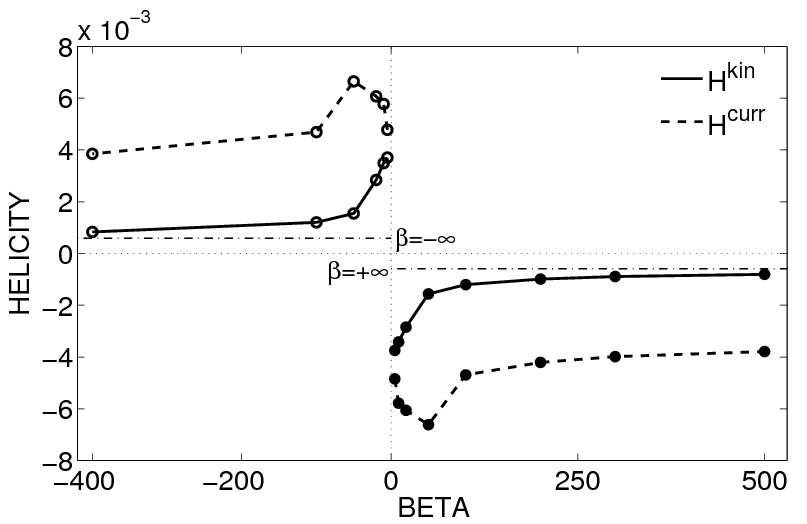}
\caption{\label{helk}
The kinetic and current helicities of the
nonaxisymmetric perturbations as functions of $\beta$. Top: for 
$\Omega>\Omega_{\rm A}$  ($\rm Ha=100$, $\rm Re=200$). Bottom: for  $\Omega_{\rm A}>\Omega$ ($\rm Ha=200$, $\rm Re=20$).  The dash-dotted lines indicate the limits
$\pm 6\cdot 10^{-4}$ of the kinetic helicity of the left and right modes in
Fig.~\ref{heltor}.}
\end{figure}

Fig.~\ref{helk}  shows the kinetic (\ref{1}) and current
(\ref{2}) helicities for the two series of runs.  For both series, both
helicities have the opposite sign as $\beta$ (see also K\"apyl\"a \& Brandenburg 2009) for
comparison). In stating this result, it is important though to specify
carefully the nature of the initial conditions used in each run.  For
$\beta=O(1)$, the basic state has a sufficiently strong handedness that it
forces the instabilities to have a particular parity as well, which as
indicated turns out to be opposite to that of the basic state.  If one then
gradually increases $\beta$, each time using the previous solution as the
new initial condition, this parity of the instabilities is preserved all
way to $\beta\to\infty$, where the basic state no longer has a handedness,
and both left and right instabilities could exist equally well, as in Fig.
\ref{heltor}.

That is, by the time one reaches $\beta=500$, say, the basic state makes
sufficiently little distinction between left and right modes that both
could exist, but because of the way we have reached $\beta=500$, we
consistently obtain the right mode.  However, suppose one does the following
experiment now: Take the right mode at $\beta=500$, swap its parity to be
left, and use that as a new initial condition for a series of runs where
$\beta$ is now gradually reduced. Eventually there comes a point where the
basic state's handedness is sufficiently great that it no longer allows the
instability to have the `wrong' parity, and the solution reverts back to
the right mode. This feature that both left and right modes are allowed for
sufficiently large $\beta$ (where the degeneracy between the two modes is
only weakly broken) but not for smaller $\beta$ (where the degeneracy is
strongly broken) is in many ways analogous to an imperfect pitchfork
bifurcation.

\section{Alpha-effect and dynamo theory}
We have also calculated the $\alpha$-effect in (\ref{3}) with the same averaging  procedure over the azimuth. Because of the complex structure of the background field it is even possible to determine parts of the tensorial structure of the $\alpha$-tensor. In particular we are interested in the signs and amplitudes of the $\alpha$-effect in both azimuthal and axial directions. According to the general rule that the azimuthal $\alpha$-effect is anticorrelated with the (kinetic) helicity we expect the azimuthal $\alpha$-effect to be positive for $\beta>0$. The expected sign of the axial $\alpha$-effect is not clear. There are theories and simulations leading to $\alpha_{\phi\phi}$ and $\alpha_{zz}$ with opposite signs (see R\"udiger \& Hollerbach 2004 for an overview). We should not be surprised to find a similar behavior in the present simulations. It also means that any dynamo with very weak differential rotation cannot be treated with a scalar $\alpha$-effect. 

\begin{figure}
\includegraphics[width=8.5cm,height=6cm]{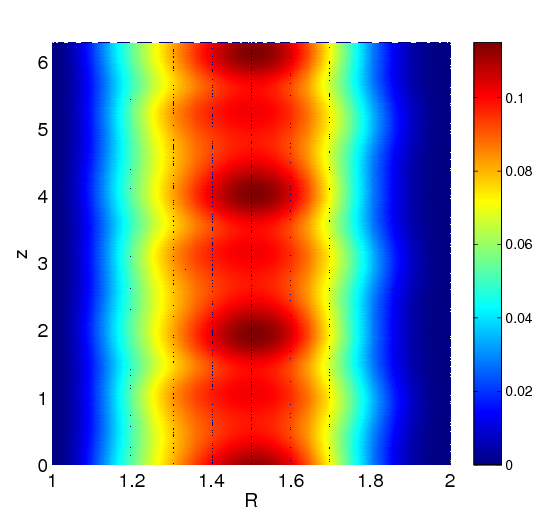}
\includegraphics[width=8.5cm,height=6cm]{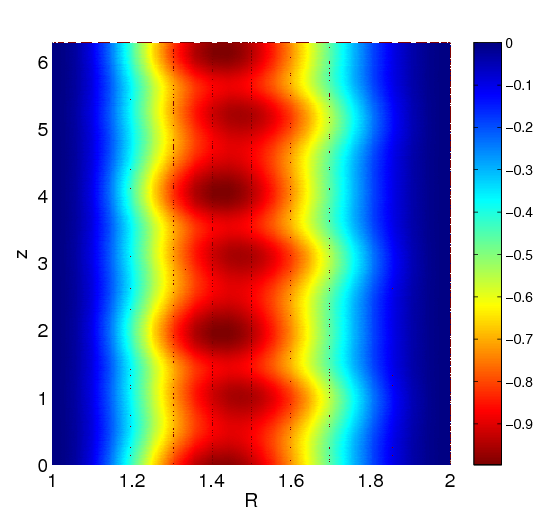}
\caption{\label{alpha1}
The $\alpha$-effect in the slowly rotating case, $\Omega<\Omega_{\rm A}$,
$\rm Re=20$, $\rm Ha=200$. The top shows $\alpha_{\phi\phi}$, the bottom
$\alpha_{zz}$. The other parameters are $\beta=3$, $\mu_\Omega=0.5$,
$\mu_B=1$, $\rm Pm=1$.}
\end{figure}

\begin{figure}
\includegraphics[width=8.5cm,height=6cm]{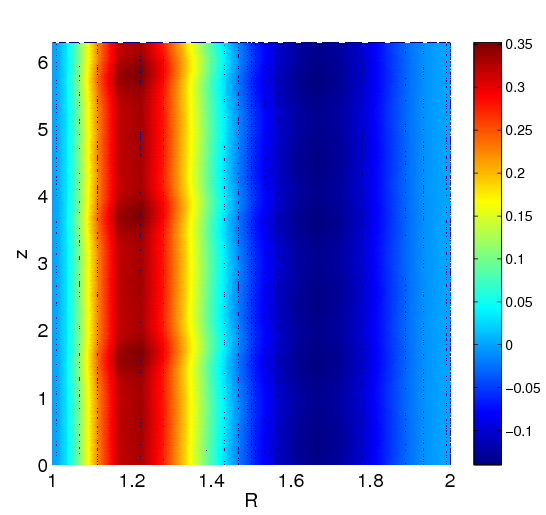}
\includegraphics[width=8.5cm,height=6cm]{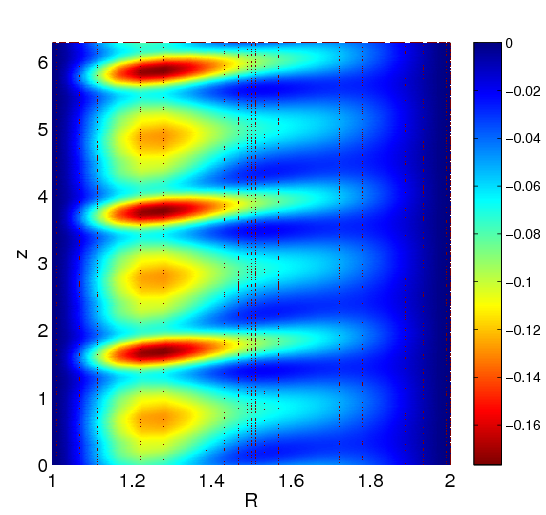}
\caption{\label{alpha2}
The $\alpha$-effect in the rapidly rotating case, $\Omega>\Omega_{\rm A}$,
$\rm Re=200$, $\rm Ha=100$. The top shows $\alpha_{\phi\phi}$, the bottom
$\alpha_{zz}$. The other parameters are $\beta=3$, $\mu_\Omega=0.5$,
$\mu_B=1$, $\rm Pm=1$.}
\end{figure}

Figs.~\ref{alpha1} and \ref{alpha2} give the results for slow and rapid
rotation. On the basis of Eq.~(\ref{3})  the dimensionless $\alpha$-effect
in the form
\begin{equation}
C_\alpha= \frac{\alpha D}{\eta}
\label{calfa}
\end{equation}
is plotted for the components $\alpha_{\phi\phi}$ and $\alpha_{zz}$.
In both cases, these two components have opposite signs, with $\alpha_{\phi\phi}
>0$ and $\alpha_{zz}<0$ almost everywhere in the meridional plane.  This
anti-correlation between the two components is also strongest in the center of
the gap, and weakest near the boundaries.  It is therefore not caused by the
boundaries.

\begin{figure}
\includegraphics[width=8.5cm,height=6cm]{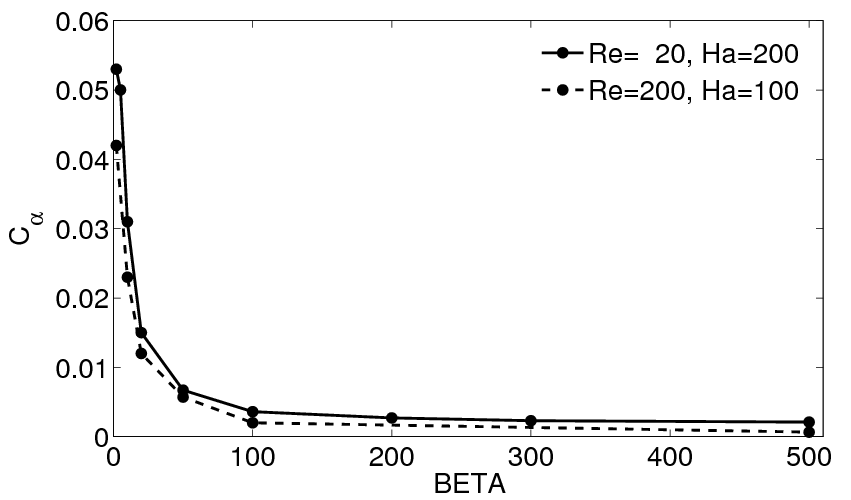}
\caption{\label{alpha}
The  dimensionless dynamo number $C_\alpha$ of the azimuthal $\alpha$-effect for 
$\Omega<\Omega_{\rm A}$ and $\Omega>\Omega_{\rm A}$.  For higher $\beta$ (smaller $B_z$) the $\alpha$-effect decreases like $C/\beta$ with $C\simeq 0.05$.
$\mu_\Omega=0.5$, $\mu_B=1$, $\rm Pm=1$.}
\end{figure}

That Figs. \ref{alpha1} and \ref{alpha2} show such similar results is
surprising, and is one of the basic results of this paper.  The influence of
rotation on $\alpha$ is evidently rather weak. The fact that -- contrary to
previous results for rotating convection -- $\alpha_{\phi\phi}$ is actually
smaller for rapid rotation than for slow rotation illustrates just how
different these magnetic-induced helicities are from some of the previous
results. Finally, Fig. \ref{alpha} shows how the amplitudes of
$\alpha_{\phi\phi}$ vary with $\beta$, being roughly inversely proportional
in both cases.

To consider some possible astrophysical implications of these results,
imagine a disk dynamo with dominant field components $B_\phi$ and $B_R$.
Dynamo waves of $\alpha\Omega$-type require for self-excitation that the
product of (\ref{calfa})  and
\begin{equation}
 C_\Omega= - \frac{D^3}{\eta} \frac{{\rm d} \Omega}{{\rm d} R}
\label{calf}
\end{equation}
exceeds a critical value of order unity, i.e. $C_\alpha C_\Omega \geq 1$.
The ratio of the amplitudes of the field components $B_\phi$ and $B_R$
follows the simple rule
\begin{equation}
\frac{|B_\phi|}{|B_R|} \simeq \sqrt{\frac{C_\Omega}{C_\alpha}}
\label{bfr}
\end{equation}
so that  dynamo excitation requires
\begin{equation}
\frac{|B_\phi|}{|B_R|} C_\alpha\geq 1 .
\label{bfr1}
\end{equation}
We know from Fig.~\ref{alpha} that $C_\alpha \simeq C/\beta$ with $C \ll 1$,
so that (\ref{bfr1}) gives, at least as an order-of-magnitude estimate,
the condition 
\begin{equation}
C > \frac{|B_R|}{|B_z|}
\label{C}
\end{equation}
for self-excitation of a dynamo with differential rotation and current-driven $\alpha$-effect. For disk dynamos $B_R$ dominates $B_z$, and for spherical
dynamos $B_R$ is comparable to $B_z$. In both cases the condition for
self-excitation becomes $C>1$, which cannot be fulfilled according to Fig.
\ref{alpha}, which suggests instead that $C\lsim 0.05$. The $\alpha$-effect
due to the current helicity of the background field appears as much too small
to allow the operation of an $\alpha\Omega$-dynamo.

Another argument concerns the growth rate of such a dynamo (if it exists at all) in relation to the very long magnetic diffusion times in radiative zones. 
Assume that for self-excitation $C_\alpha C_\Omega > 1$, then the growth rate
$\omega_{\rm gr}$  is given by
\begin{equation}
\omega_{\rm gr}= \frac{\eta}{D^2} \sqrt{C_\alpha C_\Omega} \simeq \frac{\eta}{D^2}\ C\ \frac{|B_z|}{|B_R|}.
\label{omgr}
\end{equation}
Hence, only for $C>1$ the  growth time of the dynamo would  be {\em shorter}  than the magnetic diffusion time $D^2/\eta$, which is known to be of order Gyr for the radiative interior of stars.

One can also argue as follows. The relation (\ref{omgr}) also reads
\begin{equation}
\omega_{\rm gr} \simeq \sqrt{\alpha\Omega'}
\label{omgr1}
\end{equation}
independent of the magnetic diffusivity. On the other hand, for given $C_\alpha$ (\ref{omgr1}) states
\begin{equation}
\omega_{\rm gr} \simeq \frac{1}{D}\sqrt{C_\alpha\eta \Omega},
\label{omgr2}
\end{equation}
which for the computed value $C_\alpha\simeq 0.01$ taken from Fig. \ref{alpha} and $\eta\simeq 500$ cm$^2$/s for the solar core leads to values of order $10^{-15}$ s$^{-1}$, i.e. to growth times of order 10 Myr.  As it is typical for $\alpha\Omega$-dynamos their growth times are only slightly shorter than the basic magnetic decay time.
\section{Summary}
We have shown that the current-driven instability of helical large-scale fields does produce small-scale helicity (kinetic plus current helicity) and even $\alpha$-effects, but the resulting numerical values seem to be too small for the operation of large-scale dynamos in radiative zones of early-type stars.

\end{document}